\documentclass[aps,prl,twocolumn,superscriptaddress,showpacs]{revtex4}
\usepackage{epsfig}
\usepackage{amsmath} 
\usepackage{times}

\begin{document}

[Phys. Rev. E {\bf 66}, 065103 (2002)]

\title{Range-based attack on links in scale-free networks:
are long-range links responsible for the small-world phenomenon?}

\author{Adilson E. Motter}
\affiliation{Department of Mathematics, CSSER,
Arizona State University, Tempe, Arizona 85287}

\author{Takashi Nishikawa}
\affiliation{Department of Mathematics, CSSER,
Arizona State University, Tempe, Arizona 85287}

\author{Ying-Cheng Lai}
\affiliation{Department of Mathematics, CSSER,
Arizona State University, Tempe, Arizona 85287}
\affiliation{Departments of Electrical Engineering and Physics,
Arizona State University, Tempe, Arizona 85287}

 
\begin{abstract}
The small-world phenomenon in complex networks has been identified as
being due to the presence of long-range links, i.e., links connecting
nodes that would otherwise be separated by a long node-to-node
distance. We find, surprisingly, that many scale-free networks are
more sensitive to attacks on short-range than on long-range links.
This result, besides its importance concerning network efficiency
and/or security, has the striking implication that the small-world
property of scale-free networks is mainly due to short-range links.
\end{abstract}
\maketitle

Many real networks have been identified to have an amazingly 
small average shortest path since Watts and
Strogatz (WS) \cite{WS:1998} introduced their model 
of small-world networks. This model is constructed
from a sparse regular network by rewiring a small fraction 
of links at random. Watts \cite{Watts:book}
introduced the concept of {\it range} to characterize 
different types of links: the range of a link
$l_{ij}$ connecting nodes $i$ and $j$ is the length of 
the shortest path between nodes $i$ and $j$ in the absence of $l_{ij}$
(see also Ref. \cite{Pandit:1999}).
In this sense, typically, local connections are short-range 
links but rewired connections are long-range links.
A key feature in the WS model is that it clearly identifies 
the small shortest paths observed in locally
structured, sparse networks as being due to long-range 
connections, while short-range links are responsible
for high clustering. This remarkable observation matches 
very well with the known results for the Erd\"os-R\'enyi (ER)
model of random graphs \cite{ER:1960}, where almost all 
links are long-range connections and the average shortest path
increases only logarithmically with the number $N$ of nodes \cite{Bollobas:book}.
In regular networks, on the other hand, all the links have small 
range and the average shortest path increases with a power of $N$.

The WS and ER models explain some important features 
of real networks, such as the small-world phenomenon.
However, since these models are homogeneous,
their connectivity distribution $P(k)$, where $k$ is the number 
of links connected to a node, has an exponential tail,
in contrast to the algebraic one that characterizes scale-free networks
recently discovered in a variety of real-world situations 
\cite{BA:1999,AB:2002},
\begin{equation}
P(k)\sim k^{-\gamma},
\label{scaling}
\end{equation}
where $\gamma$ is the scaling exponent. Scale-free 
networks are heterogeneous as their connectivity
can vary significantly from node to node and a considerable number 
of links can be associated with a few highly connected nodes.
Barab\'asi and Albert (BA) identified in their seminal paper \cite{BA:1999},
growth with preferential attachment as the 
universal mechanism generating the algebraic behavior (\ref{scaling}).
As most scale-free networks possess the small-world property, it has been 
{\it tacitly} assumed that long-range connections
are responsible for the small average shortest path exhibited by these networks.
In addition to the insights provided by the WS model, 
the main argument for this comes from the observation
that the removal of a link $l_{ij}$ of range $R$ increases 
the length of the shortest path between nodes $i$ and $j$
by $R-1$. The length of the shortest path between nodes 
connected by a short-range link is then robust against the removal of the link
because the second shortest path between these two nodes is still short.
But this is not true for long-range links, as they connect nodes that 
would otherwise be separated by a long shortest path.

Scale-free networks have attracted a tremendous amount of recent
interest \cite{AB:2002}. The aim of this paper is to investigate
{\it explicitly} the contribution of short-range links to the small-world
property in scale-free networks by analyzing the impact of
attacks on short-range links versus those on long-range links. 
Attack here is defined as the deliberate removal of a subset of selected links.
The importance of studying attacks on complex networks is twofold.
First, it can identify the vulnerabilities of real-world networks,
which can be used either for protection (e.g., of Internet) or for destruction
(e.g., of metabolic networks targeted by drugs).
Second, it provides guidance in designing 
more robust artificial networks (e.g., power grids).
Different aspects of attacks on complex networks have been analyzed recently
\cite{Pandit:1999,AJB:1999,callaway:2000,broder:2000,sole:2001,cohen:2001,jeong:2001}.
However, to our best knowledge, almost all the previous works consider 
attacks on nodes rather than on links, with very 
few exceptions \cite{GN:2001,HKYH:2002}.

To study {\it range}-based attacks on {\it links},
we consider the following models of scale-free networks:
(1) semirandom model \cite{NSW:2001};
(2) BA model \cite{BA:1999} and its generalization 
with aging \cite{Doro_Mendes:2000}.
In each case, we generate scale-free networks with the 
small-world property and a tunable scaling exponent.
Because of the small-world property, one might intuitively
think that these networks are much more sensitive to
attacks on long-range than on short-range links. 
Surprisingly, our analysis and numerical computation show exactly 
the opposite for many scale-free networks.
This result has an unexpected implication: short-range 
links are the vital ones for efficient communication
between nodes in these networks. Our findings are based on 
the observation that the average shortest path
is a global quantity that is mainly determined by links with large load,
where the load of a link is defined as the number of 
shortest paths passing through the link \cite{Sc_coll,goh:2001}.
For scale-free networks with exponent $\gamma$ in a finite interval
around 3, due to heterogeneity, the load 
is on average larger for links with shorter range,
making the short-range attack more destructive.
For very large values of $\gamma$, the corresponding networks become 
homogeneous and, as a result, the opposite occurs.

For a given network, our attack strategy is as follows.
We first compute the range for all the links.
We then measure the {\it efficiency} of the network as links are 
successively removed according to their ranges:
($i$) for short-range attacks, links with shorter ranges are removed first;
($ii$) for long-range attacks, links with longer ranges are removed first \cite{range}.
In both cases, the choice among links with the same range is made at random.
The efficiency is measured by the shortest paths between pairs of nodes.
The shortest path between two given nodes $i$ and $j$
is defined as the minimal number $d_{ij}$ of links necessary
to follow from one node to the other. A convenient quantity to 
characterize the efficiency is then 
\begin{equation}
E=\frac{2}{N(N-1)}\sum\frac{1}{d_{ij}},
\label{performance}
\end{equation}
where the sum is over all $N(N-1)/2$ pairs of nodes.
The network is more efficient when it has 
small shortest paths, which according to our
definition corresponds to large $E$. 
Definition (\ref{performance}) was introduced in 
Ref. \cite{LM:2001} to generalize the concept of
small world, as it applies to any network 
regardless of its connectedness.

We first consider the semirandom model as follows.
We start with $N$ nodes $\{ 1,2,\ldots,N\}$ and a list of $N$ 
integers representing their connectivities,
i.e., the number of ``half-links'' of each node: $\{k_1,k_2,\ldots,k_N\}$, 
where $k_i\leq N-1$ and $\sum_{i=1}^N k_i$ is even.
In the case of scale-free networks, this connectivity sequence 
is generated according to the algebraic distribution (\ref{scaling}).
Next, we pick up pairs of half-links at random and connect 
them to form a link and repeat this process until the
last pair is connected, prohibiting self- and repeated links. 
In order to have nontrivial networks in the limits of small and large $\gamma$,
we bound the connectivity so that $k_{min}\leq k_i\leq k_{max}$ 
for $i=1,2,\ldots,N$, where $k_{min}$ and $k_{max}$
are constant integers. For $\gamma\rightarrow\infty$, 
the network becomes a regular random graph,
which is homogeneous with all the nodes having the same connectivity $k_{min}$.
For $\gamma\rightarrow 0$, most of the links are associated with nodes with
connectivity of the order of $k_{max}$, and the network becomes densely connected.
The most interesting regime corresponds to intermediate values of $\gamma$ 
because in this case, the network is highly heterogeneous but still sparse,
having the number of links much smaller than $N(N-1)/2$.
Consider then this case.

\begin{figure}[t]
\begin{center}
\epsfig{figure=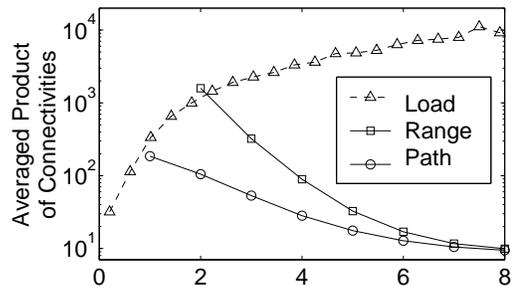,width=6.8cm}
\caption{Averaged product of connectivities as a function of the shortest path,
range, and load for $\gamma=3$, where the load is binned
and normalized by $10^4$. Each curve corresponds to the average over 10
realizations for $N=5000$, $k_{min}=3$, and $k_{max}=500$.}
\label{fig1}
\end{center}
\end{figure}

Employing the generating function formalism of Ref. \cite{NSW:2001}, 
we have derived an approximate expression for the
expected value of the shortest path between nodes 
with connectivity $k_i$ and $k_j$,
\begin{equation}
d_{ij}=\frac{\ln(Nz_1/k_ik_j)}{\ln(z_2/z_1)}+1,
\label{path}
\end{equation}
where $z_1$ and $z_2$ are the average numbers of first and second neighbors, respectively.
Accordingly, nodes with larger connectivity are on average closer 
to each other than those with smaller connectivity.
The remarkable property of Eq. (\ref{path}) is that $d_{ij}$ 
depends only on the product of the connectivities
$k_i$ and $k_j$. This relation suggests that the range is also 
correlated with the product of the connectivities \cite{pathrange}
so that short-range links tend to link together highly connected nodes,
while long-range links tend to connect nodes with very few links.
Moreover, links between nodes with large connectivities are expected 
to be passed through by a large number of shortest
paths. That is, on average, these links should possess 
a higher load \cite{HKYH:2002,load} than those connected
to nodes with fewer links. These have been confirmed numerically,
as shown in Fig. \ref{fig1} for $\gamma=3$, where we 
plot the product of connectivities averaged over
all pairs of nodes separated by a given shortest path length, or 
connected by a link with a given range or load.

Combining the above analyses for range and load, we observe that 
high load should be associated mainly with short-range links.
With the understanding that links with higher load should 
contribute more to the shortness of the paths between nodes,
this correlation between load and range implies that attacks
on short-range links are more destructive than those on long-range
links, in contrast to what one might naively think. 

\begin{figure}[t]
\begin{center}
\epsfig{figure=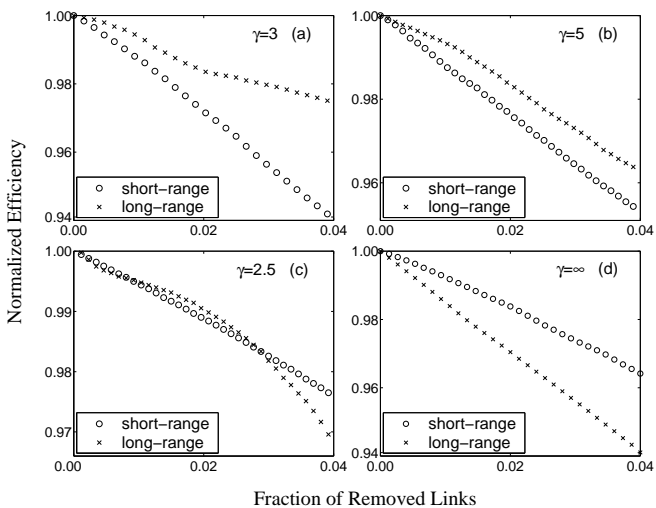,width=\linewidth}
\caption{Normalized efficiency for short- and long-range attacks
as a function of the fraction of removed links. All the parameters other than $\gamma$
are the same as in Fig. \ref{fig1}.}
\label{fig2}
\end{center}
\end{figure}

\begin{figure}[b]
\begin{center}
\epsfig{figure=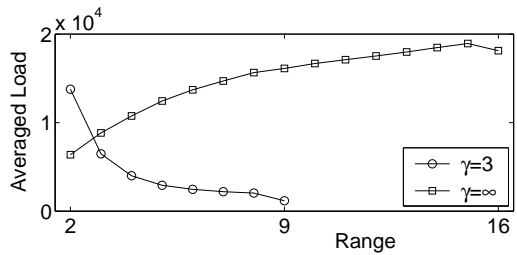,width=6.8cm}
\caption{Comparison between heterogeneous and homogeneous networks: 
averaged load as a function of the range for
$\gamma=3$ and $\gamma=\infty$. All the parameters other than $\gamma$
are the same as in Fig. \ref{fig1}.}
\label{fig3}
\end{center}
\end{figure}

Now we present numerical verification of our main result concerning
the effect of attacks on links.
In Fig. \ref{fig2}, we show the efficiency (normalized by its initial value)
for both short- and long-range attacks, for different values
of $\gamma$. Notably, short-range attacks are more destructive 
than long-range ones for intermediate values of $\gamma$, as shown in
Figs. \ref{fig2}(a) and \ref{fig2}(b) for $\gamma=3$ and $\gamma=5$,
respectively. The corresponding relation between the average
load and range, plotted in Fig. \ref{fig3} for $\gamma=3$ (open circles),
confirms that higher load on links with shorter range 
is the mechanism underlying this phenomenon.
Long-range attacks become more destructive only for networks
with sufficiently small or large values of $\gamma$.
In Figs. \ref{fig2}(c) and \ref{fig2}(d), we show the results 
for $\gamma=2.5$ and $\gamma=\infty$, respectively.
The exchange of the roles of attacks on short- and long-range
links for networks with small values of $\gamma$ is due to the
appearance of a densely connected subnetwork of nodes with large connectivity.
In this case, there are so many redundant short-range 
connections that the removal of one
will not increase the average shortest path 
by much because, for a given pair of nodes,
there are, in general, more than one path of minimal length 
which pass through {\it different} short-range links. For networks
with large values of $\gamma$, switching of the roles of short- and long-range
attacks is caused by the homogenization of the network.
In a homogeneous network, all the nodes have approximately the same connectivity.
Therefore, links with higher load are precisely those between distant nodes,
i.e., those with larger range, as shown 
in Fig. \ref{fig3} for $\gamma=\infty$ (open squares).

To demonstrate the generality of our results,
we turn next to dynamic models of scale-free networks,
where the algebraic scaling results from growth with preferential attachment,
as observed in many realistic networks \cite{BA:1999,AB:2002}.
For concreteness, we consider the BA model \cite{BA:1999} and its generalization with
aging of nodes due to Dorogovtsev and Mendes \cite{Doro_Mendes:2000}.
The model is constructed as follows.
We start at $t=0$ with $N_0$ nodes and zero links.
At each successive time step, we add a new node with $m\leq N_0$ links
so that each new link is connected to  some 
old node $i$ with probability $\Pi_i\sim \tau_i^{-\alpha}(k_i+1)$,
where $\tau_i$ is the age of the node $i$ and $k_i$ is its connectivity.
The standard BA model with scaling exponent $\gamma=3$ is recovered by taking $\alpha=0$.
In general, scale-free networks with $\gamma >2$ are generated
by choosing values of $\alpha$ in the interval $(-\infty,1]$ 
\cite{Doro_Mendes:2000}, where $\gamma$ approaches the value of
2 as $\alpha\rightarrow -\infty$ and becomes infinite as 
$\alpha\rightarrow 1$.

Most of the arguments and conclusions presented for the semirandom 
model are also valid for the growth model. In particular, 
the short-range attack is still expected to be more destructive than the
long-range one at intermediate values of $\gamma$, 
while the opposite is expected for sufficiently large $\gamma$. 
However, there is an important difference for $2<\gamma<3$.
Since new links come with new nodes,
the subnetwork of highly connected nodes must be sparse.
Accordingly, for this model, there will be no switching concerning
the effect of short- versus long-range attacks at a small value of $\gamma$.

\begin{figure}[t]
\begin{center}
\epsfig{figure=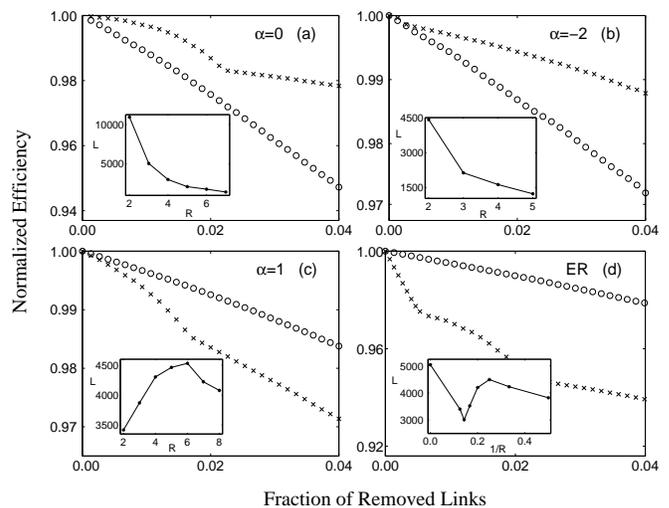,width=\linewidth}
\caption{Normalized efficiency for short-range attacks ($\circ$) and long-range attacks
(\small{$\times$}) as a function of the fraction of removed links.
Each graph corresponds to the average over 10 realizations for:
(a-c) $N=5000$, $N_0=3$, and $m=3$;
(d) $N=5000$ and $z_1=6$.
The corresponding relations between averaged loads $L$ and ranges $R$
are plotted in the insets.
Observe that in inset (d) the horizontal axis is $R^{-1}$.
}
\label{fig4}
\end{center}
\end{figure}

Our predictions are confirmed by numerical simulations,
as shown in Fig. 4 for different values of $\alpha$ ($\gamma$).
Indeed, short-range attacks are more destructive for $\alpha=0$ ($\gamma=3$) and also for
$\alpha=-2$ ($\gamma\approx 2.3$),
while long-range attacks are more destructive for $\alpha=1$
($\gamma=\infty$). In all cases, the best strategy of attack 
is consistent with the correlation between
load and range, as shown in the insets of Figs. \ref{fig4}(a)-\ref{fig4}(c).

It is instructive to compare the results for scale-free networks 
with those for homogeneous networks with
Poisson-like distribution of connectivities.
In Fig. \ref{fig4}(d), we show the efficiency 
for the ER random model \cite{comment}.
This network is  more sensitive to attacks on 
long-range links because of the strong concentration of 
load on links with range infinity (see the inset).
Incidentally, the long-range attack is also more 
destructive in the WS model \cite{WS:1998},
where the rewired connections tend to have higher load \cite{fraction}.

In summary, we have shown that for a wide interval of the scaling exponent $\gamma$, scale-free networks
are more vulnerable to short- than long-range attacks, which results from a higher concentration of load on
short-range links.
In contrast to the load-based strategies of attacks
considered in Ref. \cite{HKYH:2002}, which are based on global information, short-range attacks are {\it quasilocal}
in that, for a given range $R$, they require information only up to the $(R-1)^{th}$ neighbors \cite{local}.
Our findings have important implications that go beyond
the issue of attack itself, as they provide insights into the
structure and dynamics of scale-free networks.  
In particular, they show that short-range links are more 
important than long-range links for efficient
communication between nodes, which is the opposite to what one
might expect from other classes of small-world networks. 
For instance, in the network of sexual contacts, which is known to be scale-free \cite{sexual_contacts}, 
this means that the rapid spread of a disease may be mainly due to
short-range contacts between people with large number of partners,
in sharp contrast to its {\it homogeneous} counterpart \cite{Pandit:1999}. 

This work was supported by AFOSR under Grants Nos. F49620-01-1-0317
and F49620-98-1-0400, and by NSF under Grant No. PHY-9996454.


\begin{references}

\bibitem{WS:1998}
D.J. Watts and S.H. Strogatz, Nature (London) {\bf 393}, 440 (1998).

\bibitem{Watts:book}
D.J. Watts, {\it Small Worlds} (Princeton University Press, Princeton, 1999). 

\bibitem{Pandit:1999}
S.A. Pandit and R.E. Amritkar, Phys. Rev. E {\bf 60}, R1119 (1999).

\bibitem{ER:1960}
P. Erd\"os and A. R\'enyi, Publ. Math. Inst. Hung. Acad. Sci. {\bf 5} , 17 (1960).

\bibitem{Bollobas:book}
B. Bollob\'{a}s, {\em Random Graphs} (Academic Press, London, 1985). 

\bibitem{BA:1999}
A.-L. Barab\'{a}si and R. Albert, Science {\bf 286}, 509 (1999).

\bibitem{AB:2002}
R. Albert and A.-L. Barab\'{a}si, Rev. Mod. Phys. {\bf 74}, 47 (2002).

\bibitem{AJB:1999}
R. Albert, H. Jeong, and A.-L. Barab\'{a}si, Nature (London) {\bf 406}, 378 (2000).

\bibitem{callaway:2000}
D.S. Callaway, M.E.J. Newman, S.H. Strogatz, and D.J. Watts,
Phys. Rev. Lett. {\bf 85}, 5468 (2000).

\bibitem{broder:2000}
A. Broder, R. Kumar, F. Maghoul, P. Raghavan, S. Rajagopalan, R. Stata, A. Tomkins, and J. Wiener,
Comput. Netw. {\bf 33}, 309 (2000). 

\bibitem{sole:2001}
R.V. Sol\'e and J.M. Montoya,
Proc. R. Soc. London, Ser. B {\bf 268}, 2039 (2001).

\bibitem{cohen:2001}
R. Cohen, K. Erez, D.
ben-Avraham, and S. Havlin,
Phys. Rev. Lett. {\bf 86}, 3682 (2001).

\bibitem{jeong:2001}
H. Jeong, S.P. Mason, A.-L. Barab\'asi, and Z.N. Oltvai,
Nature (London) {\bf 411}, 41 (2001).

\bibitem{GN:2001}
M. Girvan and M.E.J. Newman, Proc. Natl. Acad. Sci. U.S.A. {\bf 99},
8271 (2002).

\bibitem{HKYH:2002}
P. Holme, B.J. Kim, C.N. Yoon, and S.K. Han,
Phys. Rev. E {\bf 65}, 056109 (2002).

\bibitem{NSW:2001}
M.E.J. Newman, S.H. Strogatz, and D.J. Watts,
Phys. Rev. E {\bf 64}, 026118 (2001).

\bibitem{Doro_Mendes:2000}
S.N. Dorogovtsev and J.F.F. Mendes,
Phys. Rev. E {\bf 62}, 1842 (2000).

\bibitem{Sc_coll}
M.E.J. Newman, Phys. Rev. E {\bf 64}, 016132 (2001).

\bibitem{goh:2001}
K.-I. Goh, B. Kahng, and D. Kim, Phys. Rev. Lett. {\bf 87}, 278701 (2001).

\bibitem{range}
We choose to sort the links according to the initial distribution of ranges,
instead of an updated distribution, because we want to address the relative importance of short-range and
long-range links for the original network. In addition, in terms of attack efficiency, updating is time consuming.

\bibitem{LM:2001}
V. Latora and M. Marchiori, Phys. Rev. Lett. {\bf 87}, 198701 (2001). 

\bibitem{pathrange}
Indeed, the range of a link
can be regarded as the length of the second shortest path between the nodes
that are connected to the link. Since we are considering the semirandom
model, for which everything other than the connectivity distribution is random,
the length of the second shortest path should also be correlated with the product of connectivities.

\bibitem{load}
For pairs of nodes connected by $n\geq 1$ shortest paths, the contribution
to the load due to each path is $1/n$.

\bibitem{comment}
In this model, we start with $N$ nodes and zero links.
Then for each pair of nodes, with probability $p$, we add a link between them.
The resulting network has on average $z_1=p(N-1)$ links per node.

\bibitem{fraction}
The same tendency displayed in Figs. 2 and 4 was observed for larger fractions of removed
links. In particular, short-range attack is still the most effective one for scale-free networks
with scaling exponent around 3.
We observe, however, that the removed fraction shown in these figures is already unrealistically
large for many practical situations.

\bibitem{local}
Reference \cite{HKYH:2002} also considers local strategies of attack.

\bibitem{sexual_contacts}
F. Liljeros, C.R. Edling, L.A.N. Amaral, H.E. Stanley, and Y. Aberg,
Nature (London) {\bf 411}, 907 (2001).


\end{references}
\end{document}